# Embedding based retrieval for long tail search queries in ecommerce

Akshay Kekuda[†]
Applied Machine Learning
Best Buy
akshay.kekuda@bestbuy.com

Yuyang Zhang
Applied Machine Learning
Best Buy
yuyang.zhang@bestbuy.com

Arun Udayashankar
Applied Machine Learning
Best Buy
arun.udayashankar@bestbuy.com

In this abstract we present a series of optimizations we performed on the two-tower model architecture [14], training and evaluation datasets to implement semantic product search at Best Buy. Search queries on bestbuy.com follow the pareto distribution whereby a minority of them account for most searches. This leaves us with a long tail of search queries that have low frequency of issuance. The queries in the long tail suffer from very spare interaction signals. Our current work focuses on building a model to serve the long tail queries. We present a series of optimizations we have done to this model to maximize conversion for the purpose of retrieval from the catalog.

The first optimization we present is using a large language model to improve the sparsity of conversion signals. The second optimization is pretraining an off-the-shelf transformer-based model on the Best Buy catalog data. The third optimization we present is on the finetuning front. We use query-to-query pairs in addition to query-to-product pairs and combining the above strategies for finetuning the model. We also demonstrate how merging the weights of these finetuned models improves the evaluation metrics. Finally, we provide a recipe for curating an evaluation dataset for continuous monitoring of model performance with human-in-the-loop evaluation. We found that adding this recall mechanism to our current term match-based recall improved conversion by 3% in an online A/B test.

## 1 User engagement-based training dataset curation

The user engagement dataset is the primary dataset that we use to build the semantic product search experience. We track different actions that a user performs on our Product List Page (PLP) and Product Detail Page (PDP) during a search session. After issuing a search, a customer can convert on a product in the following ways: PDP view of a product after click on the PLP page (pdp_view), add a product on the PLP page to cart (plp_atc), check availability of the product on the PLP page (plp_check_availabilty), add the product on PDP page to cart (pdp_atc), check availability of the product on the PDP page (pdp_check_availabilty). We take 6 months of clickstream data for our experiments and aggregate data on unique <query, product> level. This dataset comprises of both positive and negative samples. A positive sample is defined as a <query, product> pair that has at least one of the above five conversion (aka micro-conversion) signals associated with it and negative samples are those pairs without any micro-conversion signals. For our training dataset, we consider only positive samples as user judgement of non-relevance is subjective in nature.

We perform a 2-stage filtering approach on this positive dataset to reduce our training dataset size by 10x while maintaining high quality and balance across product categories. In the first stage, we only keep positive signals issued by at least 2 unique visitors which reduces false positives in the training dataset. In the second stage, we do a stratified sampling across product categories in the dataset to eliminate popularity bias.

## 2 Training data augmentation with synthetic positives

Although training of two tower models based on clickstream data is the common practice in industries [3,5,6,7,8], we find that customers at Best Buy micro convert on a small subset of products, and hence building a system based on purely engagement signals would be insufficient to understand the semantics of products that lack engagement. While collecting high quality human annotated data could be helpful [3], this is expensive in nature. Our solution to tackle this problem is to generate synthetic user queries based on product content as proxy positive signals. This is done to have a good representation of all diverse types of products sold on Best Buy in addition to just the popular ones. In addition to proxy positive signals, we also intend to capture diverse varieties of queries which are different from typical short and generic user search queries. We use Llama-13b and prompt the model to generate 10 search queries based on product title, category, description, and specifications for all products in our catalog.

## 3 Two tower model architecture

We adopt a two-tower training strategy to train our semantic search model for candidate generation. Here, the two towers correspond to the search tower and product tower, respectively. These towers are trained jointly and served independently. The search tower provides a vector for a search query and the product tower provides a vector for a product.

We first develop an in-house Best Buy BERT (B3) which is a finetuned RoBERTa base [12] model trained with the Masked Language Modeling objective on Best Buy search queries and Best Buy product fields. Taking the first five layers of B3 gives us b3-small which we use in all our experiments. On the search tower, we primarily use the preprocessed search query as the feature for the search tower. The search tower is initialized from the weights of b3-small. We use b3-small for the search tower as it needs to be used as a real time endpoint and hence latency of the search tower is crucial. On the product tower, there is flexibility in determining the modeling strategy for the product tower. A product is defined by a variety of features, and one could use some or all the fields to calculate the product vector. In our approach, we use product content such as product title, product category hierarchy, and product description as the independent product fields. We consider all features to be textual in nature. We use a mean pooling fusion layer to average product field embeddings generated using a feature encoder. Our feature encoder is again initialized from b3-small. We use the embedding of the beginning of sentence, i.e., [CLS] token to compute feature embeddings using b3-small everywhere.

## 4 Loss function

**MCCEL**: Our training dataset consists of a batch of positive samples. Inspired by [9], we used a technique called in-batch negatives to efficiently generate negative pairs during training: for each (query, product) positive pair within a training batch of size B, we take the products from the other pairs in the batch as negatives for the query of the positive pair. As a result, we will have B positive pairs and B–1 negative pairs per item in batch. We use a scaled multi-class cross entropy loss (**MCCEL**) per pair, turning this into a classification problem.

## 5 Finetuning for query-to-query similarity task

Inspired by the findings in [11], we first finetune b3-small for a query-to-query (q2q) similarity objective using the MCCEL loss. We construct a dataset of positive samples of (queryA, queryB) based on queries that have conversion on a product. We then randomly sample a maximum of 400 pairs of queries amongst queries that convert on a product. In this way, our query encoder can learn much more meaningful query embeddings that are aware of ecommerce shopping behavior. We will also be able to align synthetic queries and user queries better in this manner. We then finetune the q2q model for the query-to-product (q2p) similarity objective. We also AB test this q2q_q2p trained encoder while extending the list of product features to 13.

## 6 Model Weight Merging

Taking further inspiration from work in [10], we see three types of finetuning of b3-small in our exercise: q2q, q2p, q2q_q2p. These models have the same architecture but have been finetuned for different tasks. We do a weighted sum of these models as 0.2*q2q+0.4*q2p+0.4*q2q_q2p to be used as a feature encoder for inference. We assign a higher weight to models trained with the q2p objective over q2q objective to improve relevant product recall.

## 7 Curation of evaluation dataset

While the idea of withholding a subset of the user engagement dataset for evaluation is straightforward, it has a few key issues, namely skewed distribution, sparsity, position bias, and presentation bias. To overcome these limitations, we have adopted random sampling and human relevance judgements, drawing inspirations from the seminal works of Text REtrieval Conference (TREC) Ad hoc test collections [1] and the TREC Million Query Track [2], especially the ideas of "pooling" [1] and "wide and shallow" sampling [2]. In particular, 1) we sample from user search history in proportion to the logarithm of search count to create a more balanced search query sample; 2) we retrieve products for the sampled search queries from multiple sources, including production and experimental retrieval systems, to form the product sampling pool; 3) for each query, we sample from the product pool roughly in inverse proportion to the mean rank of the product across all sources; 4) we contract with an annotation service platform to have annotators judge the <query, product> relevance on a 4-level scale, i.e., Irrelevant, Acceptable, Good, and Excellent. Besides, to maintain the quality of the dataset, we resample new examples for labeling every month and keep adding results from new promising models to the product sampling pool.

## 8 Results

We observed that the 2-stage filtering approach of the training dataset improves recall@200 on the evaluation dataset by 0.2% over just the 1$^{st}$ stage filtering **[Section 1]**. On adding synthetic positive signals to this dataset, we see an incremental lift of 0.7% in recall@200 **[Section 2]**. Usage of the Siamese b3-small trained with 2-stage filtered dataset and synthetic positives provided an additional 1.15% lift in recall@200 **[Section 3]**. Finetuning of Siamese b3-small with the q2q similarity objective followed by finetuning for q2p similarity objective further improves recall@200 by 0.99%. This approach also improves recall@200 by 0.8% over top200 products generated using the full all-mpnet-base-v2 [13] as encoders for search and product towers. We also see that the full all-mpnet-base-v2 has the least recall@25 compared to all our custom trained models indicating better ranking of top25 products of our custom trained models which is crucial for long tail queries that are not subject to engagement-based re-rankers. Additional product features used further helped improve recall@200 by 1.5% **[Section 5]**. We finally observe a significant lift of 2.45% in recall@200 when using the merged model over the q2q_q2p model as feature encoder for inference **[Section 6].**

## 9 System Integration for Production

Some key challenges and our solutions in integrating semantic search as an additional recall method into our existing system are as follows. 1) Our existing system is heavily dependent on and oriented around Solr, which is used as not only as the inverted index but also the search results processing engine for filtering, ranking, and user filters/facets generation, making the integration with an independent vector database extremely difficult. Therefore, we have used Solr also as our vector database for indexing product embeddings and performing approximate nearest neighbor search. 2) Due to administrative reasons, our infrastructure is distributed across two cloud providers, with the existing system stack on one platform while the new ML stack, i.e., the query embedding service, on the other, which adds non-trivial network. To mitigate the impact, our engineering team has developed a caching layer to avoid redundant cross-cloud requests to the query embedding service. We hope our solution will provide a valuable reference for organizations who are faced with similar challenges.

# REFERENCES

[1] D. K. Harman, "The TREC ad hoc experiments," in TREC: Experiment and Evaluation in Information Retrieval (Digital Libraries and Electronic Publishing), pp. 79–98, MIT Press, 2005.

[2] B. Carterette and R. Jones, "Evaluating search engines by modeling the relationship between relevance and clicks," Advances in Neural Information Processing Systems, vol. 20, pp. 217–224, 2008.

[3] Zhang, Han, et al. "Towards personalized and semantic retrieval: An end-to-end solution for e-commerce search via embedding learning." Proceedings of the 43rd International ACM SIGIR Conference on Research and Development in Information Retrieval. 2020.

[4] Li, Sen, et al. "Embedding-based product retrieval in taobao search." Proceedings of the 27th ACM SIGKDD Conference on Knowledge Discovery & Data Mining. 2021.

[5] Liu, Yiding, et al. "Pre-trained language model for web-scale retrieval in baidu search." Proceedings of the 27th ACM SIGKDD Conference on Knowledge Discovery & Data Mining. 2021.

[6] Alessandro Magnani, Feng Liu, Suthee Chaidaroon, Sachin Yadav, Praveen Reddy Suram, Ajit Puthenputhussery, Sijie Chen, Min Xie, Anirudh Kashi, Tony Lee, and Ciya Liao. 2022. Semantic Retrieval at Walmart. In Proceedings of the 28th ACM SIGKDD Conference on Knowledge Discovery and Data Mining (KDD '22). Association for Computing Machinery, New York, NY, USA, 3495–3503. https://doi.org/10.1145/3534678.3539164

[7] Nigam, Priyanka, et al. "Semantic product search." Proceedings of the 25th ACM SIGKDD International Conference on Knowledge Discovery & Data Mining. 2019.

[8] Kocián, Matěj, et al. "Siamese bert-based model for web search relevance ranking evaluated on a new czech dataset." Proceedings of the AAAI Conference on Artificial Intelligence. Vol. 36. No. 11. 2022.

[9] Yiqun Liu, Kaushik Rangadurai, Yunzhong He, Siddarth Malreddy, Xunlong Gui, Xiaoyi Liu, and Fedor Borisyuk. 2021. Que2Search: Fast and Accurate Query and Document Understanding for Search at Facebook. In Proceedings of the 27th ACM SIGKDD Conference on Knowledge Discovery & Data Mining (KDD '21). Association for Computing Machinery, New York, NY, USA, 3376–3384. https://doi.org/10.1145/3447548.3467127

[10] Wortsman, Mitchell, et al. "Model soups: averaging weights of multiple fine-tuned models improves accuracy without increasing inference time." International conference on machine learning. PMLR, 2022.

[11] Liu, Zheng, et al. "Towards Generalizable Semantic Product Search by Text Similarity Pre-training on Search Click Logs." arXiv preprint arXiv:2204.05231 (2022).

[12] Liu, Yinhan, et al. "Roberta: A robustly optimized bert pretraining approach." arXiv preprint arXiv:1907.11692 (2019).

[13] Reimers, Nils, and Iryna Gurevych. "Sentence-bert: Sentence embeddings using siamese bert-networks." arXiv preprint arXiv:1908.10084 (2019).

[14] Po-Sen Huang, Xiaodong He, Jianfeng Gao, Li Deng, Alex Acero, and Larry Heck. 2013. Learning deep structured semantic models for web search using clickthrough data. In Proceedings of the 22nd ACM international conference on Information & Knowledge Management (CIKM '13). Association for Computing Machinery, New York, NY, USA, 2333–2338. https://doi.org/10.1145/2505515.2505665

# AUTHOR BIOS

**Akshay Kekuda** obtained his master's degree in computer science from The Ohio State University with a research focus in NLP. Prior to this, he was a Research Engineer at Center for Development of Telematics for 3 years. During his time at CDOT, he worked as an embedded software engineer in addition working on reinforcement learning solutions for networks. After that he worked as Graduate Research Associate at SafeAuto Insurance for 2 years building multi-turn conversation understanding capabilities for SafeAuto Insurance call centers. Currently, he works as a Machine Learning Scientist at Best Buy building end to end solutions for AI powered search experiences at Best Buy.

**Yuyang Zhang** obtained his master's degree in data science from the University of Minnesota. He has been working in data analytics and machine learning for the past 8 years. He started as a data analyst at JD.COM in China and then joined Best Buy upon graduation from graduate school. At Best Buy, he has focused on developing end to end ML solutions for Recommendations and Search.

**Arun Udayashankar** obtained his PhD in neuroscience from the Goethe University in Frankfurt, Germany. He has been working in machine learning for the past 10 years. He started as a postdoctoral associate at Oregon Health and Science University. Following that worked as a data scientist at Citibank. After that he worked at Amazon in consumer payments, catalog team and Alexa. The focus of his work at Amazon was recommendations and NLP. At Best Buy as Associate Director of machine learning he leads a team focusing on search, recommendations, and Generative AI.